\documentclass[twocolumn,prb,showpacs,preprintnumbers,superscriptaddress,amsmath,amssymb,10pt,aps]{revtex4-1}
\usepackage{graphicx}
\usepackage{dcolumn}
\usepackage{color}
\usepackage{bm}
\usepackage[hidelinks]{hyperref}
\hypersetup{
    colorlinks,
    citecolor=blue,
    filecolor=blue,
    linkcolor=blue,
    urlcolor=blue
}

\newcommand{\im}{\textrm{i}}

\DeclareMathOperator{\e}{e}
\DeclareMathOperator{\C}{\mathcal{C}}
\DeclareMathOperator{\T}{\mathcal{T}}

\DeclareMathOperator{\Tr}{Tr}

\begin{document}

\title{A bulk-boundary correspondence for dynamical phase transitions\\ in one-dimensional topological insulators and superconductors}

\author{N. Sedlmayr}
\email{ndsedlmayr@gmail.com}
\affiliation{Department of Physics and Medical Engineering, Rzesz\'ow University of Technology, al.~Powsta\'nc\'ow Warszawy 6, 35-959 Rzesz\'ow, Poland}
\affiliation{Department of Physics and Astronomy, Michigan State University, East Lansing, Michigan 48824, USA}
\author{P. Jaeger}
\affiliation{Department of Physics, University of Wuppertal, D-42097 Wuppertal, Germany}
\affiliation{Department of Physics and Astronomy, University of Manitoba, Winnipeg R3T 2N2, Canada}
\author{M. Maiti}
\affiliation{Bogoliubov Laboratory of Theoretical Physics, Joint Institute for Nuclear Research, 141980 Dubna, Moscow Region, Russia}
\author{J. Sirker}
\affiliation{Department of Physics and Astronomy, University of Manitoba, Winnipeg R3T 2N2, Canada}

\date{\today}

\begin{abstract}
We study the Loschmidt echo for quenches in open one-dimensional
lattice models with symmetry protected topological phases. For
quenches where dynamical quantum phase transitions do occur we find
that cusps in the bulk return rate at critical times $t_c$ are
associated with sudden changes in the boundary contribution. For our
main example, the Su-Schrieffer-Heeger model, we show that these
sudden changes are related to the periodical appearance of two
eigenvalues close to zero in the dynamical Loschmidt matrix. We
demonstrate, furthermore, that the structure of the Loschmidt spectrum
is linked to the periodic creation of long-range entanglement between
the edges of the system.
\end{abstract}


\maketitle

\section{Introduction}
Two important concepts in modern condensed matter physics are
topological and dynamical phase transitions. These two ideas differ
from the traditional phenomenology of phase transitions which can be
traced back to Landau.\cite{Landau1980} Key to the Landau
classification of phase transitions is the concept of order parameters
indicative of symmetries broken across phase boundaries. Contrarily,
topological phase transitions separate regions with the same
symmetries but with different topological properties of the ground
state. Dynamical phase transitions---rather than focusing on the
non-analyticities which occur in the derivatives of the free
energy---concern non-analyticities which occur in dynamical
quantities after a perturbation of the system.

The definition of a dynamical phase transition which we want to follow
here is based on the Loschmidt echo\cite{Heyl2013} 
\begin{equation}
\label{LE1}
L(t)= \langle \Psi_0 | \e^{-iH_1t} | \Psi_0\rangle.
\end{equation}
Here $|\Psi_0\rangle$ is the initial state of the system before the
quench and $H_1$ the time-independent Hamiltonian which induces the
unitary time evolution of the system. The Loschmidt echo can be viewed
as a `partition function' with fixed boundaries. Similar to the
canonical partition function\cite{Fisher1965,Bena2005} there are
`Fisher zeros' for complex times $t$ with the Loschmidt echo vanishing
if the Fisher zeros are real or approach the real axis in the
thermodynamic limit. For the Ising model in a transverse field the
authors of Ref.~\onlinecite{Heyl2013} have shown that the Loschmidt
echo becomes zero at real critical times $t_c$ only for quenches
across the critical point, i.e. in cases where the initial state is
the ground state of the Hamiltonian on one side of the transition
while the time-evolving Hamiltonian belongs to the other phase. In
this case there is therefore a direct connection between the
equilibrium and the dynamical phase transition. In recent years,
dynamical phase transitions have also been studied in a number of
other
models.\cite{Mitra2012a,Karrasch2013,Heyl2014,Andraschko2014,Sharma2016,Karrasch2017,Gomez-Leon2017,Sedlmayr2017a}
Contrary to the transverse Ising model it has been found that, in
general, there is no connection between equilibrium and dynamical
phase transitions: crossing an equilibrium phase transition does not
necessarily lead to zeros at real times in the Loschmidt echo while
such zeros can also occur for quenches within the same
phase.\cite{Andraschko2014,Vajna2014} A special case are quenches in
Gaussian models with topological order. Here it has been shown that,
under certain conditions, a quench across a topological phase
transition is guaranteed to lead to dynamical phase transitions while
the opposite is not true.\cite{Vajna2015} The phase of the Loschmidt
echo can then be used to define a dynamical topological order
parameter which changes at critical times $t_c$.\cite{Budich2016}

A natural question to ask which we want to address in this manuscript
is then: is there a dynamical analogue to the bulk boundary
correspondence of equilibrium topological phase transitions? Our paper is organized as follows. In
Sec.~\ref{Models} we introduce the class of models we will discuss. In
Sec.~\ref{LE} we review known results for the Loschmidt echo and the
return rate in the periodic case. We then present results of
numerical calculations of the return rate for open systems in
Sec.~\ref{Numerics}. To understand the origins of the observed sudden
changes of the boundary contribution to the return rate at dynamical
phase transitions we investigate the dynamical entanglement structure
in Sec.~\ref{Entangle}. Our results are summarized in
Sec.~\ref{Concl}.

\section{Models}
\label{Models}
We focus here on one-dimensional (1D) models with symmetry protected
topological (SPT) phases. Following the ten-fold way symmetry
classification\cite{Schnyder2008} we have three symmetry classes with
ground states labeled by a $\mathbb{Z}$ topological invariant AIII,
BDI, and CII; and two labeled by a $\mathbb{Z}_2$ topological
invariant D, and DIII. The unitary particle-hole operator $\C$ for
BDI, D, and DIII obeys $\{\C,H\}=0$ with $H$ the Hamiltonian of the
system and $\C^2=1$. The other two symmetries we require are the time
reversal symmetries $\mathcal{T}_\pm$ satisfying
$\mathcal{T}_\pm^2=\pm1$. They must also anticommute with $\C$. BDI
has additionally $[\mathcal{T}_+,H]=0$ and a $\mathbb{Z}$ invariant in
1D. DIII has additionally $[\mathcal{T}_-,H]=0$ and a $\mathbb{Z}_2$
invariant in 1D. As $\mathcal{T}_-$ is the time reversal symmetry of
the electrons, DIII has Kramer's pairs. D, with a $\mathbb{Z}_2$
invariant in 1D, has no additional symmetry beyond particle-hole. If
both $\mathcal{T}_\pm$ symmetries are present the system is best
thought of as in BDI with an additional TR symmetry protecting the
Kramer's pairs and this will have a $\mathbb{Z}$ invariant in
1D.\cite{Schnyder2008,Sedlmayr2015a}

We will use examples principally in the BDI class, which possesses
both a unitary `time reversal symmetry', and a particle hole
symmetry. It therefore also possesses chiral or sublattice
symmetry. The topological superconductors in which Majorana bound
states are sought all belong to either BDI, D, or
DIII.\cite{Kitaev2001,Lutchyn2010,Oreg2010,Alicea2011,Mourik2012,Wong2012}

For concreteness we consider 1D Hamiltonians, which after a Fourier
transform on a periodic lattice, are of the form
\begin{equation}
\label{1dham}
H=\sum_{ k}\Psi^\dagger_{ k}\mathcal{H}( k)\Psi_{ k} \textrm{ with } \mathcal{H}(k)=\mathbf{d}_k \cdot {\bm\tau}\,,
\end{equation}
where ${\bm\tau}$ is the vector of Pauli matrices acting in some
subspace and $\Psi_{ k}$ are the appropriate operators for that
subspace. This will be particle-hole space for examples such as the
Kitaev chain, which is a topological superconductor, or a unit-cell
subspace for the Su-Schrieffer-Heeger (SSH) chain. These two examples
will be those we focus on and we will introduce them in more detail
below. In general $\mathbf{d}_k=(d^x_k,d^y_k,d^z_k)$, and
diagonalizing $\mathbf{d}_k
\cdot {\bm\tau}$ one can find $\mathbf{\tilde d}_k \cdot
{\bm\tilde\tau}$ with $\mathbf{\tilde d}_k=(0,0,\epsilon_k)$ and such
Hamiltonians have pairs of eigenenergies $\pm \epsilon_k$, a result of
the particle-hole symmetry of the Hamiltonians we consider.

\subsection{SSH model}
\label{SSH}
The first example we will consider is the SSH model with open or
periodic boundary conditions,
\begin{equation}
\label{ssh_model}
H=-J\sum_{j}\left[(1+\delta\e^{i\pi j})c^\dagger_jc_{j+1}+\textrm{H.c.}\right]\,.
\end{equation}
Here $J$ is the nearest-neighbor hopping amplitude, $\delta$ is the
dimerization, and $c^\dagger_j$ is the creation operator on site
$j$. For periodic boundary conditions the sum is taken up to site $N$
and $c_{N+1}=c_N$ while the sum is taken up to $N-1$ for open boundary
conditions. The main reason to consider this specific model first, is
that an exact solution for open boundary conditions
exists\cite{Shin1997,Sirker2014a} which depends on a set of parameters
determined by non-linear equations. As a result, numerically accurate
data for very large open systems can be easily obtained. The system is
topologically non-trivial for $\delta>0$. The particle-hole symmetry
is then $\psi_j\to\im\psi_j^\dagger$ and $T_+$ is
$\psi_j\to(-1)^j\psi_j^\dagger$. Note that the phase of $\C$ must be
fixed such that $\{\C,\T_+\}=0$.

For periodic boundary conditions the Hamiltonian can be easily diagonalized. Firstly after a Fourier transform and a convenient rotation,
\begin{equation}
\Psi^\dagger_k=\sqrt{\frac{2}{N}}\sum_{j=1}^{N/2}\e^{\im 2kj}\begin{pmatrix}
1&0\\0&\e^{\im 2k}
\end{pmatrix}\underbrace{\begin{pmatrix}
c_{2j-1}\\
c_{2j}
\end{pmatrix}}_{=\Psi_j}\,,
\end{equation}
the Hamiltonian takes the form \eqref{1dham} with
\begin{equation}
\label{d_SSH}
\mathbf{d}_k=\begin{pmatrix}
-2J\cos[k],2J\delta\sin[k],0
\end{pmatrix}\,,
\end{equation}
which can be readily diagonalized. The momenta are $k=2\pi n/N$ with
$n=1,2,\ldots N/2$. The particle-hole symmetry is now
$\C=\e^{\im\pi/2}{\bm \tau}^x\hat K$ and $\T_+=\hat K$, where $\hat K$ is the complex conjugation operator. 

Although this model has a $\mathbb{Z}$ winding number, the values are
nevertheless confined to be either 0 or 1. Extensions of this model
in the same symmetry class but with a higher winding number are however
possible.\cite{Rice1982}

\subsection{Long-range Kitaev chain}
\label{Kitaev}
As our second example we will consider the Kitaev chain of $M$
sites with long-range hopping terms,\cite{Kitaev2001} 
\begin{eqnarray}\label{khlr}
H&=&\sum_{i,j}\Psi^\dagger_{i}\left(
\Delta_{|i-j|}\im{\bm\tau}^y-J_{|i-j|}{\bm\tau}^z\right)\Psi_{j+1}+\textrm{H.c.}\nonumber\\&&-\mu\sum_{j}\Psi^\dagger_{j}{\bm\tau}^z\Psi_{j}\,,
\end{eqnarray}
with open or periodic boundary conditions. For the periodic case we
have $\Psi_{M+1}=\Psi_{1}$.  The operators in particle-hole space are
given by $\Psi^\dagger_{j}=(c^\dagger_{j},c_{j})$, and $c_{
j}^{(\dagger)}$ annihilates (creates) a spinless fermionic particle at
a site $j$. In this case we have again $\C=\e^{\im\pi/2}{\bm
\tau}^x\hat K$, $\T_+=\hat K$, and a Fourier transform brings the
Hamiltonian into the form of Eq.~\eqref{1dham} where
$\Psi^\dagger_k=(c^\dagger_k,c_{-k})$ and
\begin{equation}
\label{d_Kit}
\mathbf{d}_k=\sum_{m=1}^3 \begin{pmatrix}
-2J_m\cos[mk]-\mu/3,2\Delta_m\sin[mk],0
\end{pmatrix}\,.
\end{equation}
The long-range hopping has been truncated here at a distance of 3
sites and we define $\vec J=(J_1,J_2,J_3)$ and
$\vec\Delta=(\Delta_1,\Delta_2,\Delta_3)$ . Note that contrary to the
SSH model phases with higher winding numbers in $\mathbb{Z}$ exist,
allowing for a more general investigation of quenches between
topological phases with different invariants, and therefore also
different numbers of boundary states.  The momenta are $k=2\pi n/M$
with $n=1,2,\ldots M$ and the total system size is $N=2M$.

\section{The Loschmidt echo and return rate}
\label{LE}
In this section we will define the quantities studied throughout the following,
and review results for periodic boundary conditions. 
The initial state $|\Psi_0\rangle$ in Eq.~\eqref{LE1} is the many-body
ground state of the initial Hamiltonian $H_0$ before the quench. The
unitary time evolution is then determined by the Hamiltonian
$H_1$. The Loschmidt echo in a translationally invariant system of the
form of Eq.~\eqref{1dham} can be easily calculated\cite{Quan2006} as
the momentum $k$ remains a good quantum number during the quench. One
finds
\begin{equation}\label{pbcloschmidt}
L(t)=\prod_k\left[\cos(\epsilon^1_kt)+\im\hat{\mathbf{d}}^0_k\cdot\hat{\mathbf{d}}^1_k\sin(\epsilon^1_kt)\right]\,,
\end{equation}
with
$\hat{\mathbf{d}}^{0,1}_k=\mathbf{d}^{0,1}_k/\sqrt{\mathbf{d}^{0,1}_k\cdot\mathbf{d}^{0,1}_k}$
and $\mathbf{d}^{0,1}_k$ being the parameter vector in the
Hamiltonian \eqref{1dham} before and after the quench respectively. The
product in $k$ is over all filled states of the lower band.

More generally, for any free fermion system the Loschmidt echo can
always be obtained from the single-particle correlation matrix defined
by
\begin{equation}
\label{Cij}
\C_{ij}=\langle\Psi_0|\Psi^\dagger_i\Psi_j|\Psi_0\rangle \, .
\end{equation}
Here $i$ and $j$ run over all lattice sites. The Loschmidt echo in
terms of the correlation matrix is given
by\cite{Levitov1996,Klich2003,Rossini2007}
\begin{equation}
\label{rle}
L(t)=\det\mathbf{M}\equiv\det\left[1-\mathbf{\C}+\mathbf{\C}\e^{\im {\bm H}_1t}\right]\,.
\end{equation}
Here ${\bm H}_1$ is the Hamiltonian matrix written in the same basis
as $\C$. We will use \eqref{rle} to calculate $L(t)$ in the open
boundary case and call $\mathbf{M}$ the {\it Loschmidt matrix} in the
following. Eq.~\eqref{pbcloschmidt} is easily recovered for periodic
boundary conditions by transforming to momentum space and by using the
eigenbasis of the initial (momentum resolved) Hamiltonian.

In a many-body system we expect, in general, an orthogonality
catastrophe: the overlap between the initial state and the states in
the time evolution will become exponentially small in system size
$N$. It is therefore useful to define the return rate as
\begin{equation}
\label{return}
l(t)=-\frac{1}{N}\ln|L(t)|\,.
\end{equation}
The non-analytic points of the return rate are determined by the zeros
of the Loschmidt echo.\cite{Heyl2013} Fisher zeros in the complex
plane occur in the translationally invariant case at
times\cite{Vajna2015}
\begin{equation}
\label{tn1}
t_n(k)=\frac{\pi}{\epsilon^1_k}\left(n+\frac{1}{2}\right)+\frac{i}{\epsilon^1_k}\tanh^{-1}\left(\hat{\mathbf{d}}^0_k\cdot\hat{\mathbf{d}}^1_k\right)\,
\end{equation}
wit $n$ being an integer. These zeros lie on the real axis and
therefore give rise to non-analytic behavior for the return rate at critical
times
\begin{equation}
\label{tn2}
t_n=\frac{\pi}{2\epsilon^1_{k^*}}\left(2n-1\right)\,, \textrm{ where  }n\in\mathbb{Z}\,,
\end{equation}
if a critical momentum $k^*$ exists with
\begin{equation}
\label{tn3}
\hat{\mathbf{d}}^0_{k^*}\cdot\hat{\mathbf{d}}^1_{k^*}=0 \, .
\end{equation}
This is the condition for the vanishing of the imaginary part in
Eq.~\eqref{tn1}. We introduce the critical time scale
$t_c=\pi/(2\epsilon^1_{k^*})$. Where multiple critical times exist, we
take the smallest critical time to be the timescale $t_c$.

\section{Boundary contributions to the return rate}
\label{Numerics}
In this study we are interested in the boundary contributions to the
return rate \eqref{return} for systems with open boundaries. In the
large $N$ limit we can expand the return rate as
\begin{equation}
\label{bbreturn}
l(t)\sim l_0(t)+\frac{l_{B}(t)}{N}\,.
\end{equation}
Here $l_0$ is the bulk contribution which is equivalent to the return
rate in the thermodynamic limit for periodic boundary
conditions. $l_B$ is the boundary contribution which contains information about the topologically protected edge states, as we will demonstrate  in the following.

\subsection{Finite-size scaling}
\label{scaling}
The most straightforward approach to find $l_B(t)$ is a numerical
calculation of the correlation matrix \eqref{Cij} followed by a finite
size scaling analysis of the return rate, Eq.~\eqref{bbreturn}. We
will discuss the results of such an approach here for both our
examples, the SSH and the long-range Kitaev chain.

\subsubsection{SSH chain}
\begin{figure}[!ht]
\includegraphics*[width=0.99\linewidth]{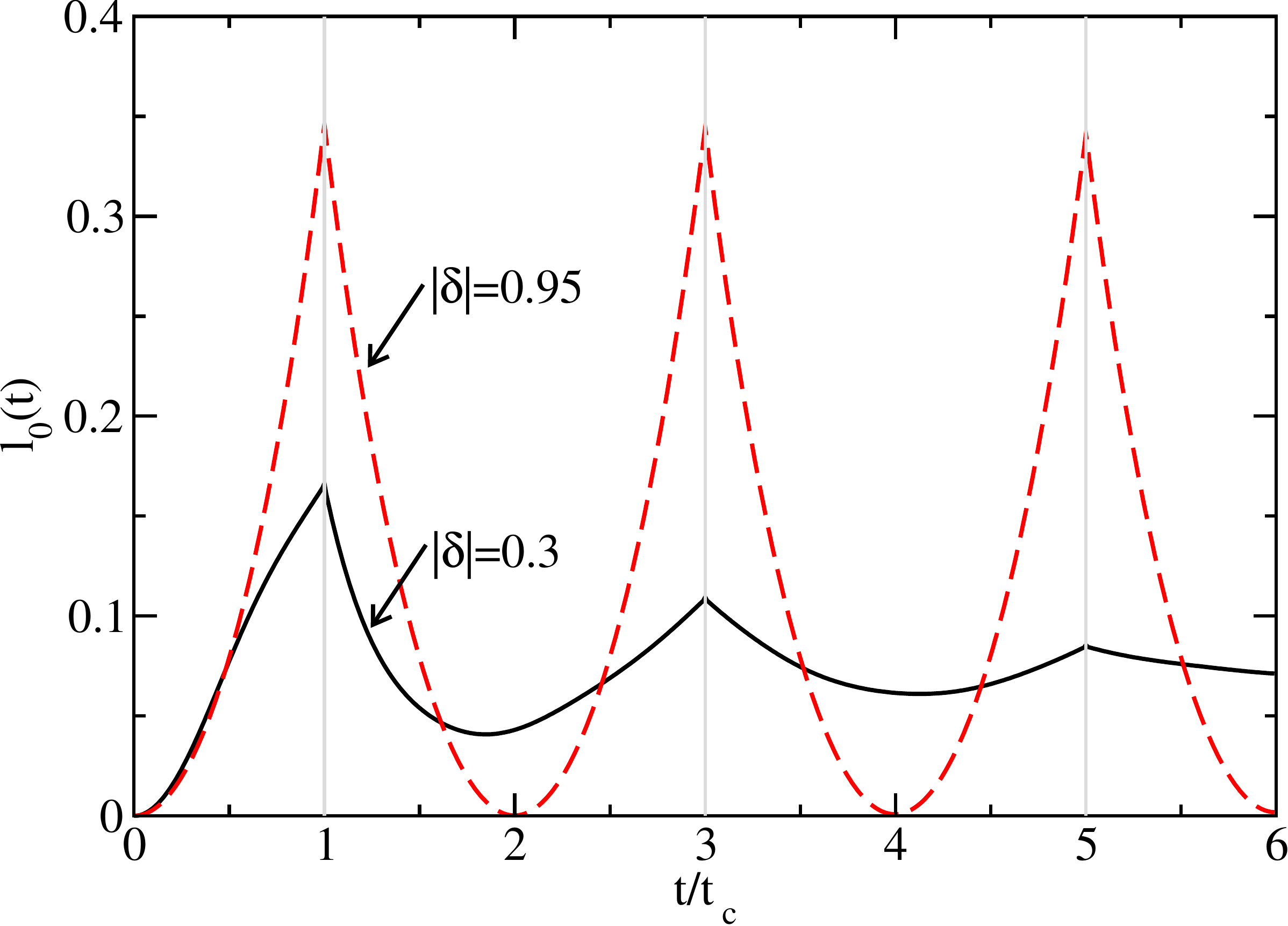}
\caption{Bulk contributions for symmetric quenches $\delta\to -\delta$ in the SSH model for $\delta=0.3$ and $\delta=0.95$.
}
\label{Fig1}
\end{figure}

We start with the SSH chain where the semi-analytical solution for
open boundary conditions allows one to obtain highly accurate results
for very large systems. We consider quenches between the topologically
ordered phase with edge states ($\delta>0$) and the topologically
trivial phase without edge states ($\delta<0$) in the half-filled
case. If we perform a symmetric quench $\delta\to -\delta$ then the
direction of the quench does not matter for the bulk contribution
$l_0(t)$ shown in Fig.~\ref{Fig1} as is obvious from
Eq.~\eqref{pbcloschmidt}.
In both examples cusps in the bulk return rate are present at the
critical times determined by Eqs.~(\ref{tn2}, \ref{tn3}).

\begin{figure}[!ht]
\includegraphics*[width=0.99\linewidth]{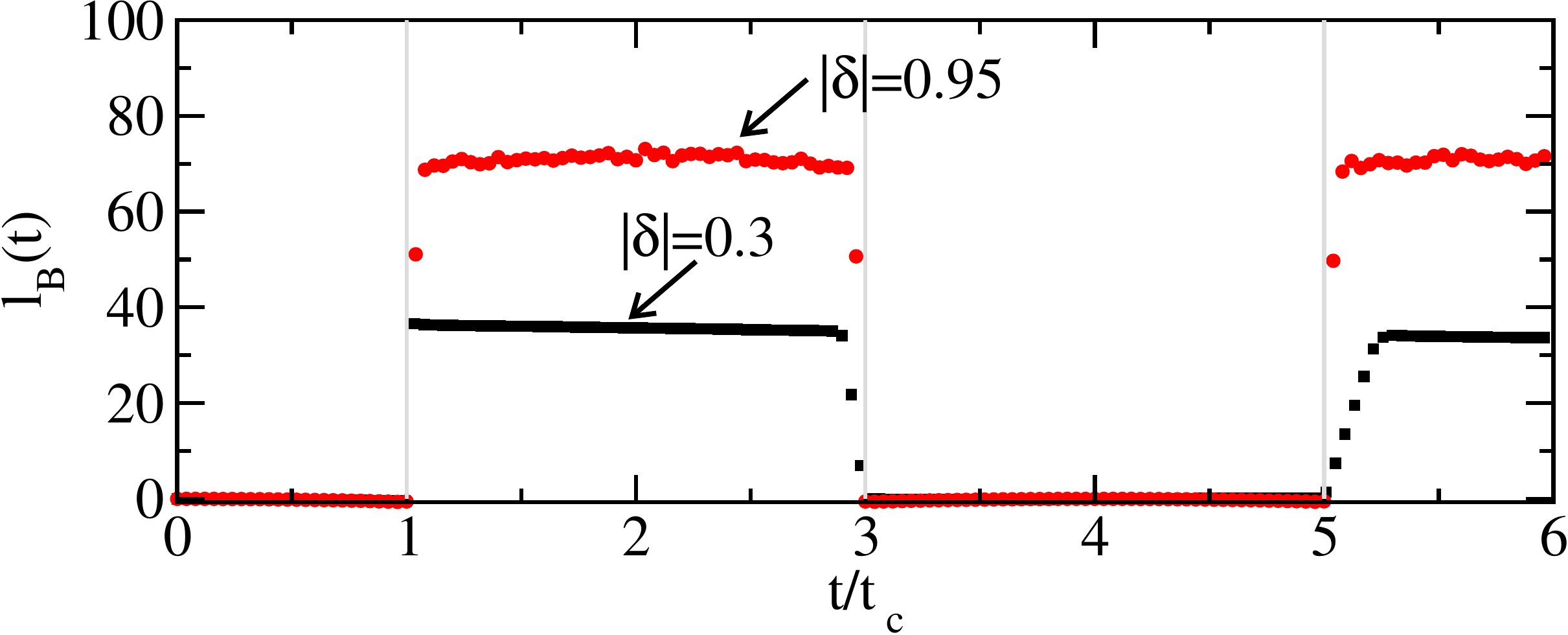}
\caption{Boundary contribution to the return rate $l_B(t)$ for symmetric quenches from 
the trivial into the topological phase. $l_B(t)$ is extracted from a
$1/N$ scaling analysis of chains of up to $N=2200$ sites.}
\label{Fig2}
\end{figure}

\begin{figure}[!ht]
\includegraphics*[width=0.99\linewidth]{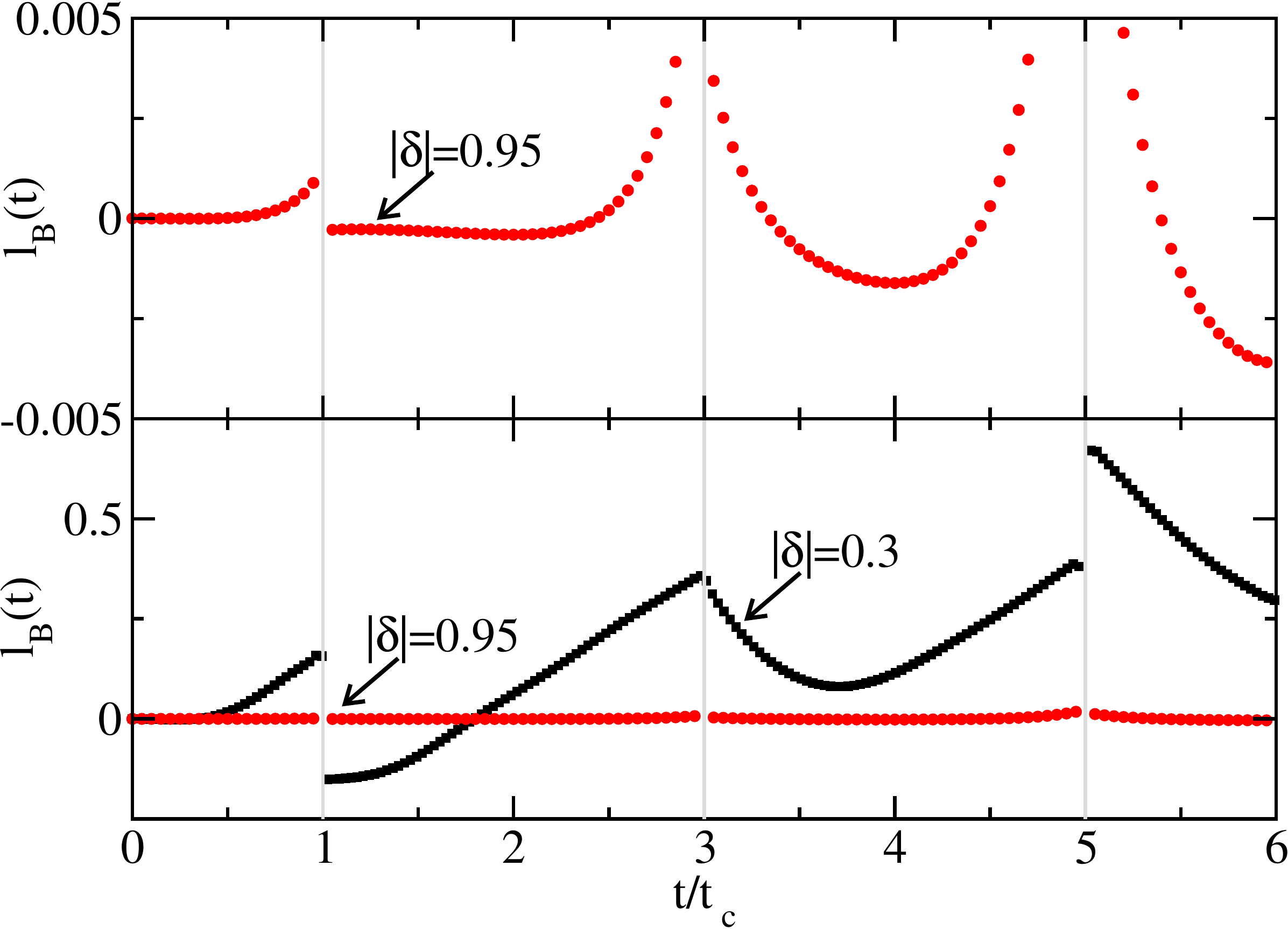}
\caption{Boundary contribution to the return rate $l_B(t)$ for symmetric quenches from 
the topological into the trivial phase. $l_B(t)$ is extracted from a
$1/N$ scaling analysis of chains of up to $N=2200$ sites. Note the
different scale of $l_B(t)$ compared to the quenches in
Fig.~\ref{Fig2}. The top panel shows a close-up of $l_B(t)$ for the quench from $\delta=0.95$ to $\delta=-0.95$.}
\label{Fig3}
\end{figure}

\begin{figure}[!ht]
\includegraphics*[width=0.99\linewidth]{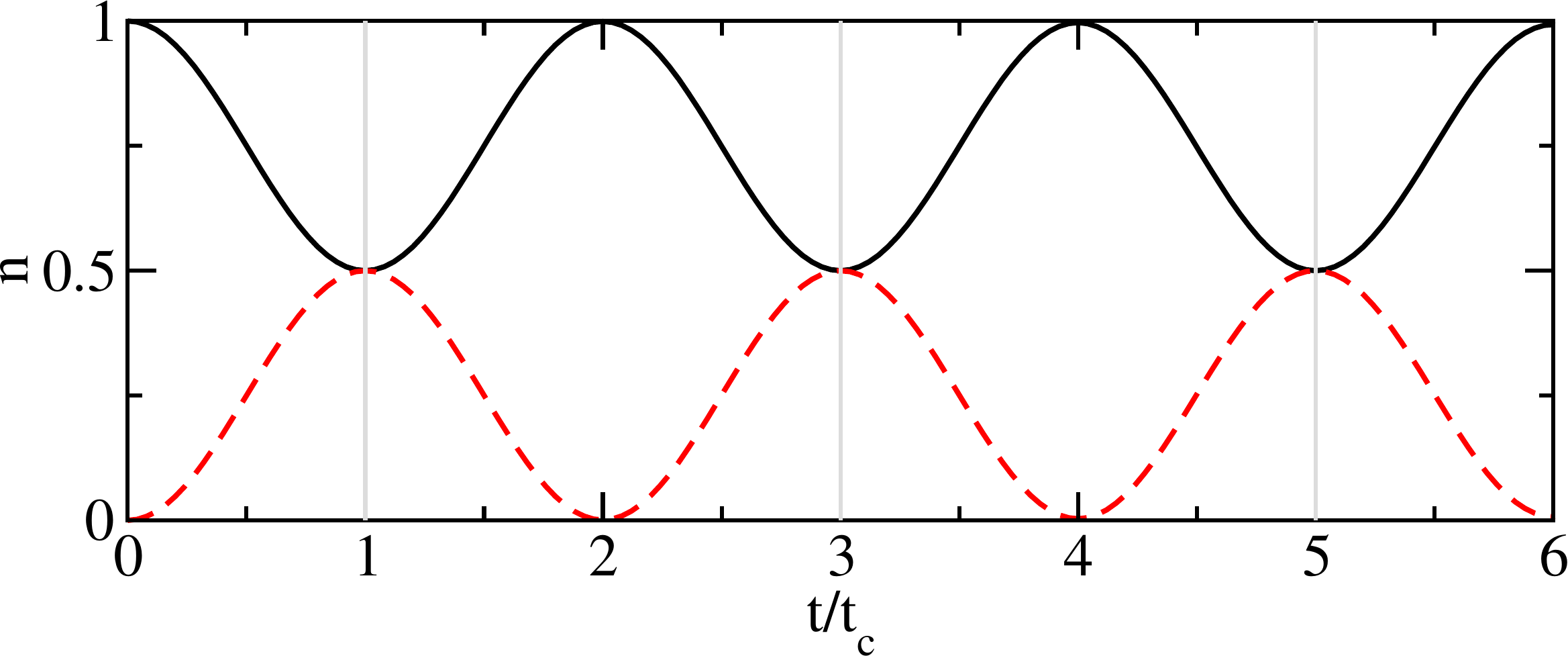}
\caption{Occupation of the edge modes for a symmetric quench at $|\delta|=0.95$ from the topological to the trivial phase for a system of size $N=80$.}
\label{Fig4}
\end{figure}

The direction of the quench does, however, strongly affect the
boundary contribution $l_B(t)$. For a quench from the trivial into the
topological phase we find a boundary contribution $l_B(t)$ which shows
large jumps at the critical times, see Fig.~\ref{Fig2}.
For a quench from the topological into the trivial phase we also
observe jumps in $l_B(t)$ at $t_c$ as shown in Fig.~\ref{Fig3},
however, the jumps are more than two orders of magnitude smaller than
for the quench in the opposite direction.
The data in the two figures clearly point to a bulk-boundary
correspondence: At the same critical times where the bulk contribution
shows cusps and the bulk dynamical winding number
changes,\cite{Budich2016} the boundary contribution also shows
discontinuities. The dependence on the direction of the quench
furthermore suggests, that the boundary contribution $l_B(t)$ is
strongly affected by the presence or absence of symmetry protected
edge states in the final Hamiltonian. That the boundary contribution
is directly related to the edge states can be seen from the
time-dependent occupation of these states, see Fig.~\ref{Fig4}.
For a system with $N/2$ spinless fermions in the topological phase,
one of the edge modes is filled at time $t=0$ while the other one is
empty. At the critical times $t_c$, where the cusps in $l_0(t)$ occur,
both edge modes are approximately half-filled. In the remainder of the
paper we will investigate the relation between the edge modes and the
singularities in the Loschmidt echo in more detail.

\subsubsection{Long-range Kitaev chain}
\begin{figure}[!ht]
\includegraphics*[width=0.99\linewidth]{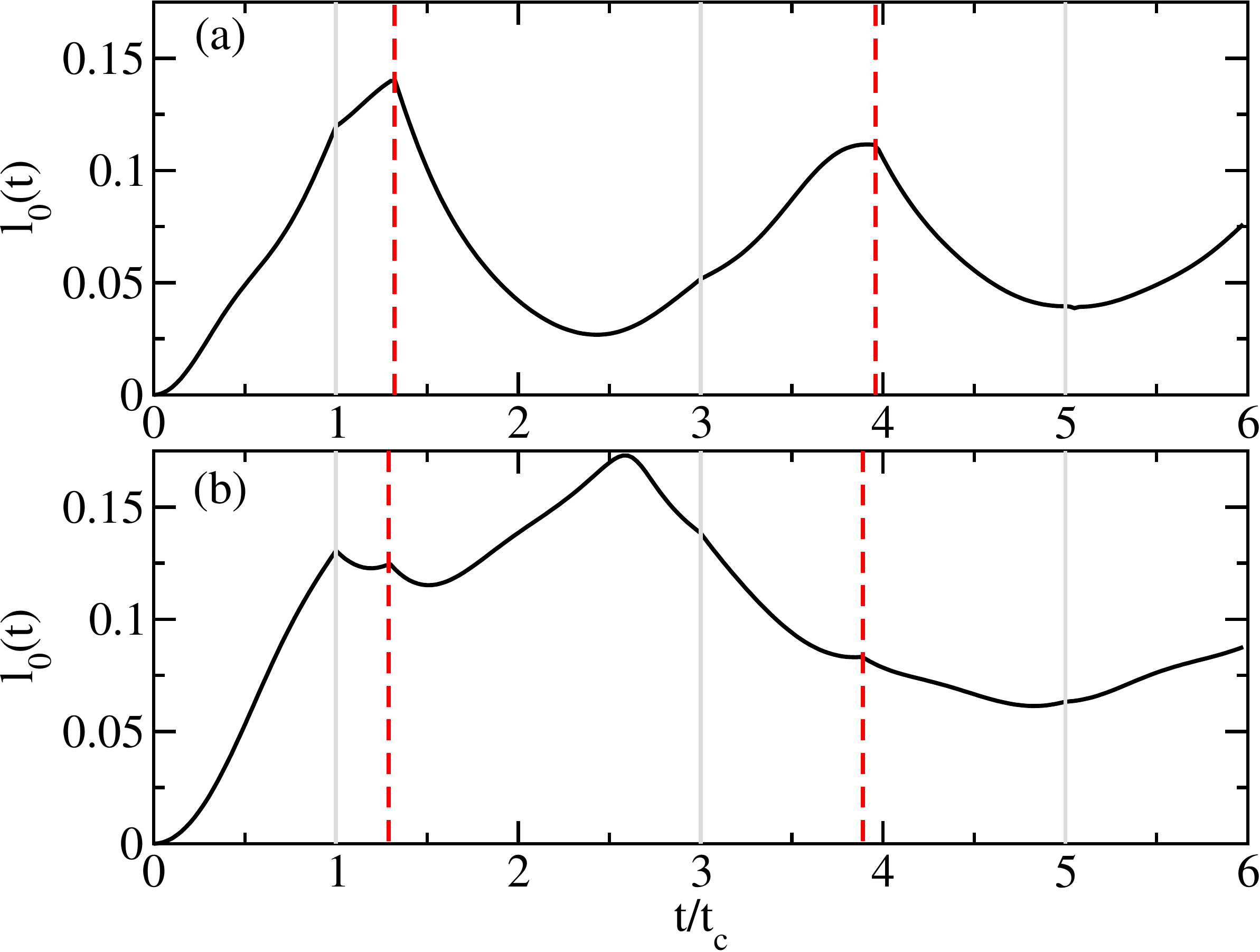}
\caption{Bulk return rate for quenches in the 
long-range Kitaev model from (a) $\nu_{\textrm{w}}=1$ to $\nu_{\textrm{w}}=3$ and (b) $\nu_{\textrm{w}}=1$ to $\nu_{\textrm{w}}=-1$. The gray and dashed red lines show the positions of the critical times.}
\label{Fig5}
\end{figure}

Before doing so, we will first present numerical results for $l_B(t)$
for the other model system we study here, the long-range Kitaev
chain. Contrary to the SSH model, this chain has different topological
phases characterized by an integer winding number
$\nu_{\textrm{w}}$. Dynamical phase transitions are expected for any
quench between phases with different winding numbers. In
Fig.~\ref{Fig5} the bulk return rate is shown for two examples. 
In the case shown in Fig.~\ref{Fig5}(a) the quench is from
$\nu_{\textrm{w}}=1$ with $\vec J=(1,-2,2)$, $\mu=2$, and
$\vec\Delta=(1.3,-0.6,0.6)$ to $\nu_{\textrm{w}}=3$ with $\vec
J=(1,-2,2)$, $\mu=0.1$, and $\vec\Delta=(0.45,-0.9,1.35)$. Two
critical momenta exist leading to two distinct critical times at which
DPT's occur. In Fig.~\ref{Fig5}(b) a quench from $\nu_{\textrm{w}}=1$
with $\vec J=(1,-2,-2)$, $\mu=3$, and $\vec\Delta=(1.3,-0.6,0.6)$ to
$\nu_{\textrm{w}}=-1$ with $\vec J=(1,-2,-2)$, $\mu=3$, and
$\vec\Delta=-(1.3,-0.6,0.6)$ is considered. Cusps in the return rate
are again clearly visible, demonstrating that DPT's also occur for
quenches where only the sign of the winding number changes.

\begin{figure}[!ht]
\includegraphics*[width=0.99\linewidth]{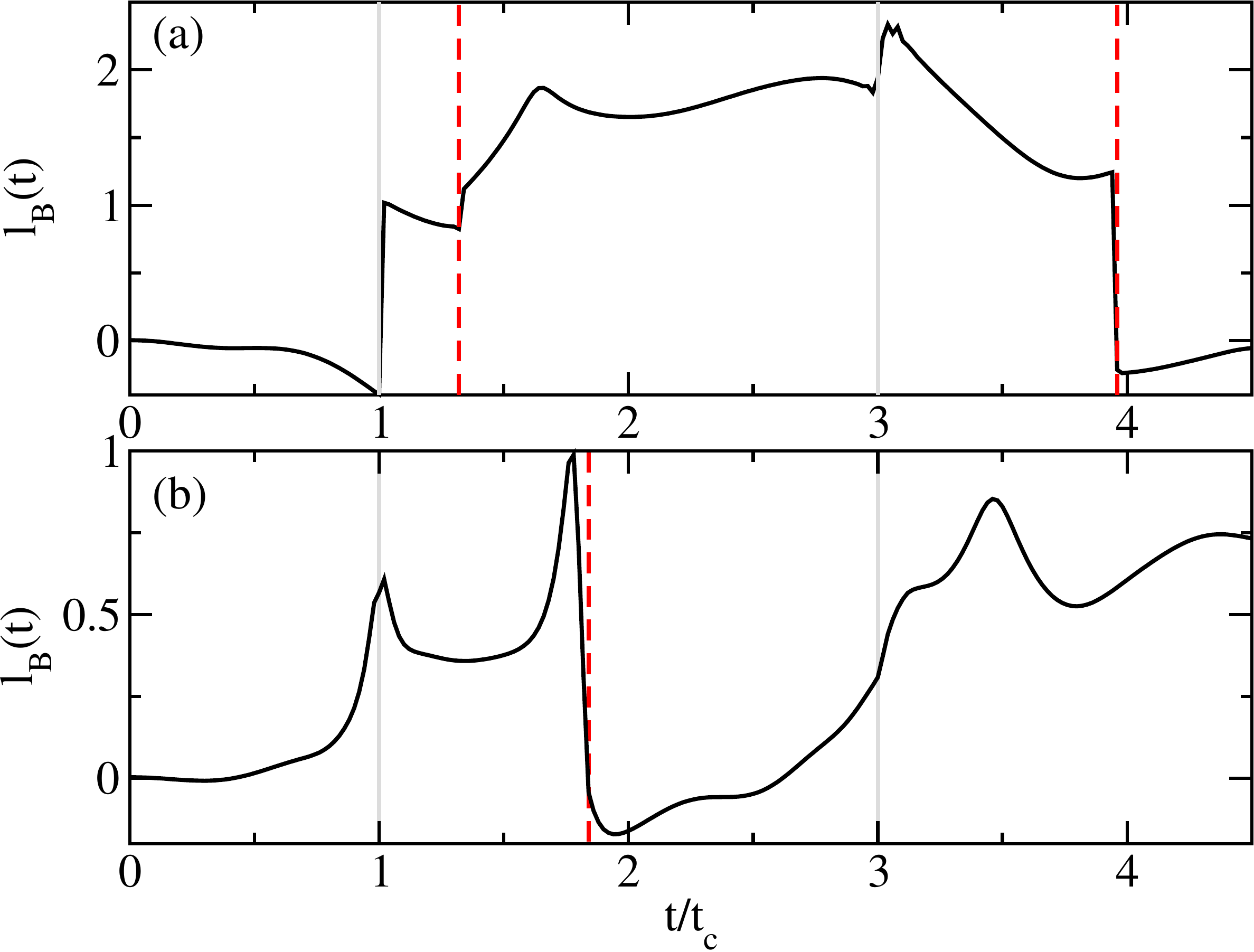}
\caption{Boundary contribution $l_B(t)$ in the long-range Kitaev model for (a)
the same quench as in Fig.~\ref{Fig5}(a) from $\nu_{\textrm{w}}=1$ to
$\nu_{\textrm{w}}=3$, and (b) the reverse quench from
$\nu_{\textrm{w}}=3$ to $\nu_{\textrm{w}}=1$. $l_B(t)$ is extracted from a
$1/N$ scaling analysis of chains of up to $N=1600$. The gray and dashed red
lines show the positions of the critical times.}
\label{Fig6}
\end{figure}

By a finite size scaling analysis of chains up to $N=1600$, we
have also extracted the boundary contribution. As an example we show
$l_B(t)$ for the same quench as in Fig.~\ref{Fig5}(a) in
Fig.~\ref{Fig6}(a), and for the reverse quench in Fig.~\ref{Fig6}(b). Note that contrary to the SSH model an analytic
solution for the eigensystem is not known so that we cannot increase
the system size till a clear scaling also for times very close to
$t_c$ emerges. Finite-size corrections are still present near
DPT's. Furthermore, the edge states have energies which are
exponentially close to zero so that an exact diagonalization in
multi-precision is required.
While the obtained data are therefore not as accurate as for the SSH
chain, we nevertheless observe a behavior which is qualitatively
similar. At the critical times $t_c$ the boundary contribution shows
jumps. However, here the jumps are of similar magnitude for both
quench directions.

\subsection{Loschmidt eigenvalues}
In order to understand the origin of the discontinuous boundary
contribution we next investigate the spectrum of the dynamical
Loschmidt matrix $\mathbf{M}$ defined in Eq.~\eqref{rle}.

\begin{figure}[!ht]
\includegraphics*[width=0.99\linewidth]{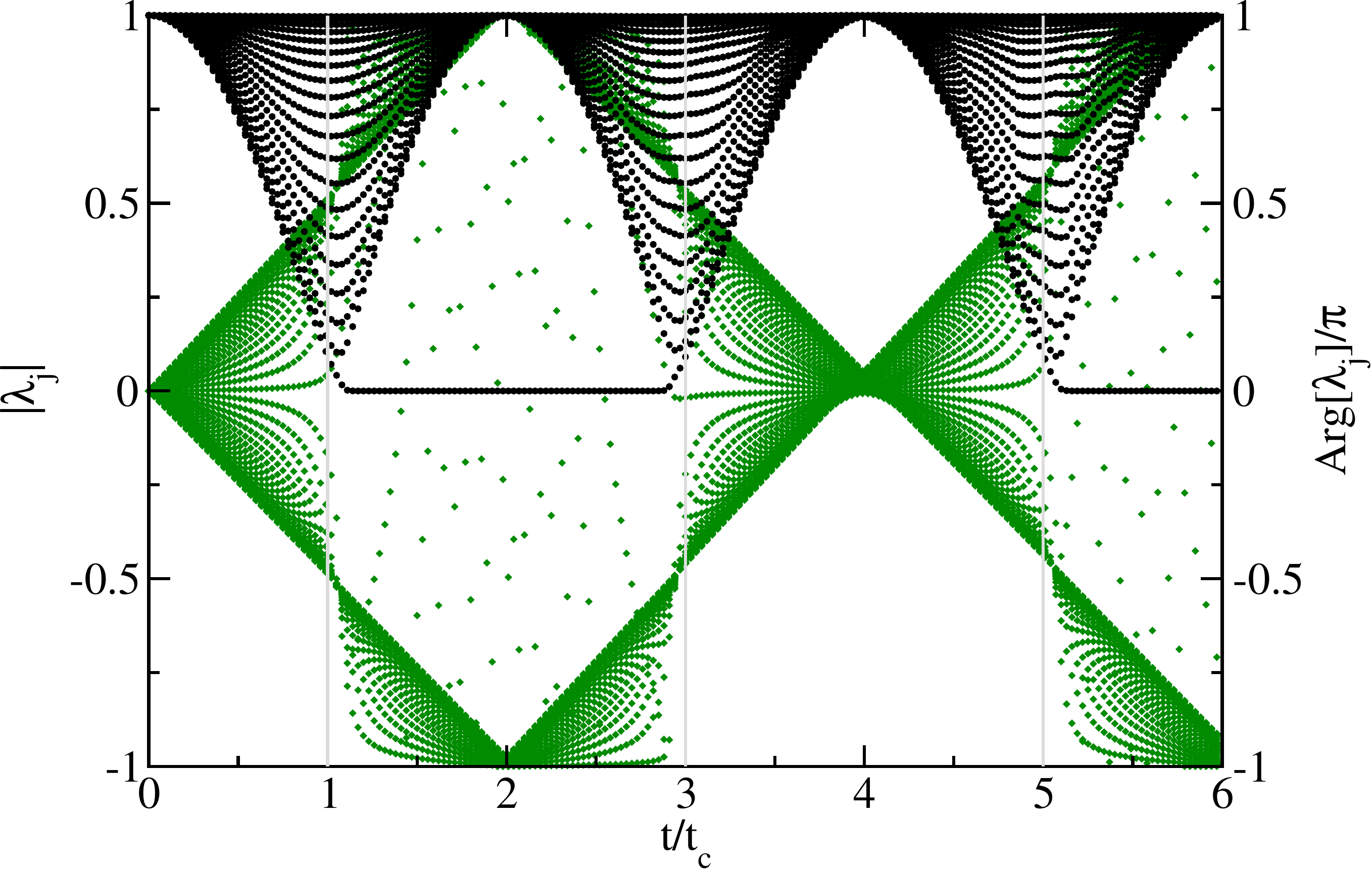}
\caption{Absolute value [black circles] and argument [green diamonds] of the eigenvalues of the matrix $\mathbf{M}$ for a symmetric quench at $|\delta|=0.95$ from the trivial into the topological phase for a system of size $N=80$.}
\label{Fig7}
\end{figure}

We concentrate first on symmetric quenches in the SSH model for large
dimerizations $\delta$ where the structure of the spectrum is
particularly simple. In Fig.~\ref{Fig7} the spectrum of the matrix
$\mathbf{M}$ for a quench from the trivial into the topological phase
is shown.
At the first critical time $t_c$ there are two eigenvalues
$\lambda_{1,2}$ whose absolute values become exponentially small in
system size. Between $t_c$ and $3t_c$ these eigenvalues stay close to
zero and this structure repeats itself in time. At $t_c$, the argument
of the eigenvalues also changes abruptly from zero to $\pm\pi$. These
two eigenvalues are related to the edges of the system as a comparison
to the case with periodic boundary shows where they are absent.

\begin{figure}[!ht]
\includegraphics*[width=0.99\linewidth]{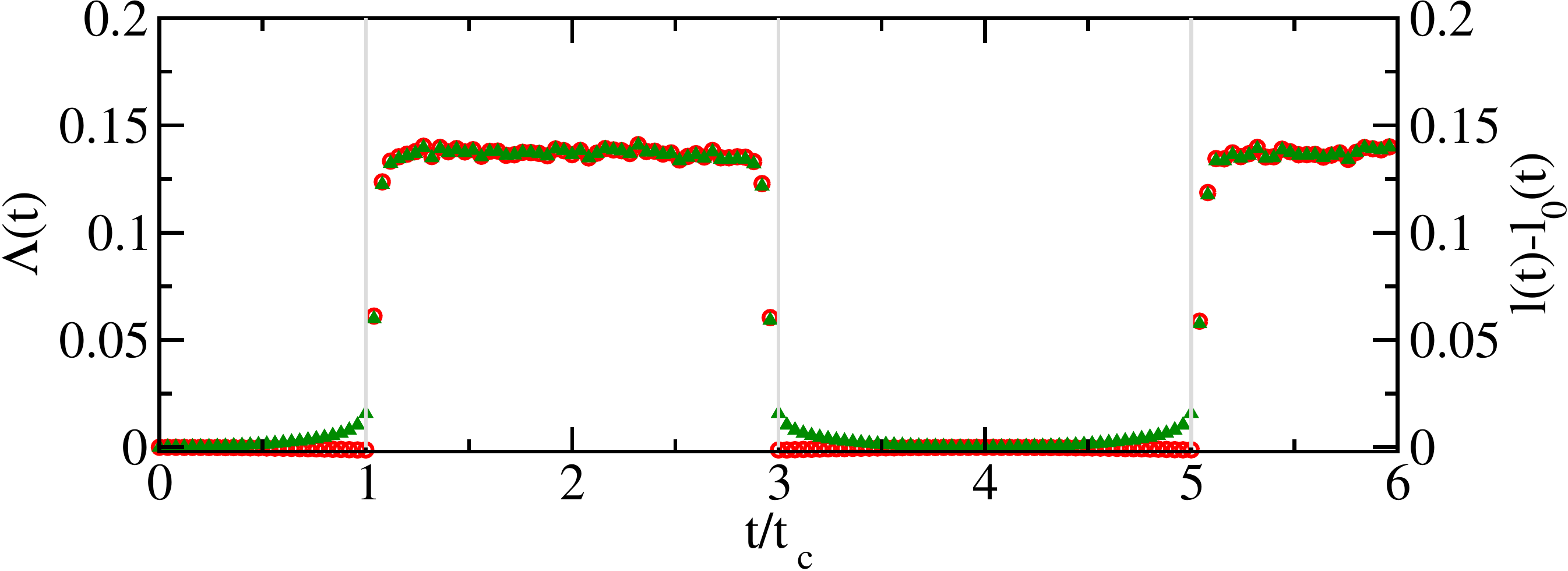}
\caption{The boundary contribution to the return rate $l(t)-l_0(t)$ [red circles] compared to $\Lambda$ [green triangles], the contribution to the return rate from
the two eigenvalues which are close to zero for $t\in [t_c,3t_c]$, see Fig.~\ref{Fig7}. Shown for a system of
size $N=500$.}
\label{Fig8}
\end{figure}

The return rate in terms of the eigenvalues $\lambda_i$ of the matrix
$\mathbf{M}$ is given by
\begin{equation}
\label{EVs}
l(t)=-\frac{1}{N}\sum_{j=1}^N \ln |\lambda_j| \,.
\end{equation}
We can isolate the contribution to $l(t)$ from $\lambda_{1,2}$, the two eigenvalues which periodically approach
zero:
\begin{equation}
\label{2EVs}
\Lambda\equiv-\frac{1}{N}\ln |\lambda_1\lambda_2|\,.
\end{equation}
At a particular system size $N$, $\Lambda$ reproduces the same  behavior as the finite size boundary contribution which we define as $l(t)-l_0(t)$, see
Fig.~\ref{Fig8}. It is the principle 
 source of the large boundary contribution for $t\in[t_c,3t_c]$.

At a DPT the system thus not only changes back and forth between the
trivial phase with dynamical winding number $\nu=0$ and the
topological one with $\nu=1$\cite{Budich2016} there is also---at the
same time---a transition in the edge degrees of freedom. The presence
or absence of a pair of zero eigenvalues of the dynamical matrix
$\mathbf{M}$ can serve as an equivalent order parameter to the bulk
winding number, establishing a concrete bulk-boundary correspondence
in this case.

\begin{figure}[!ht]
\includegraphics*[width=0.99\linewidth]{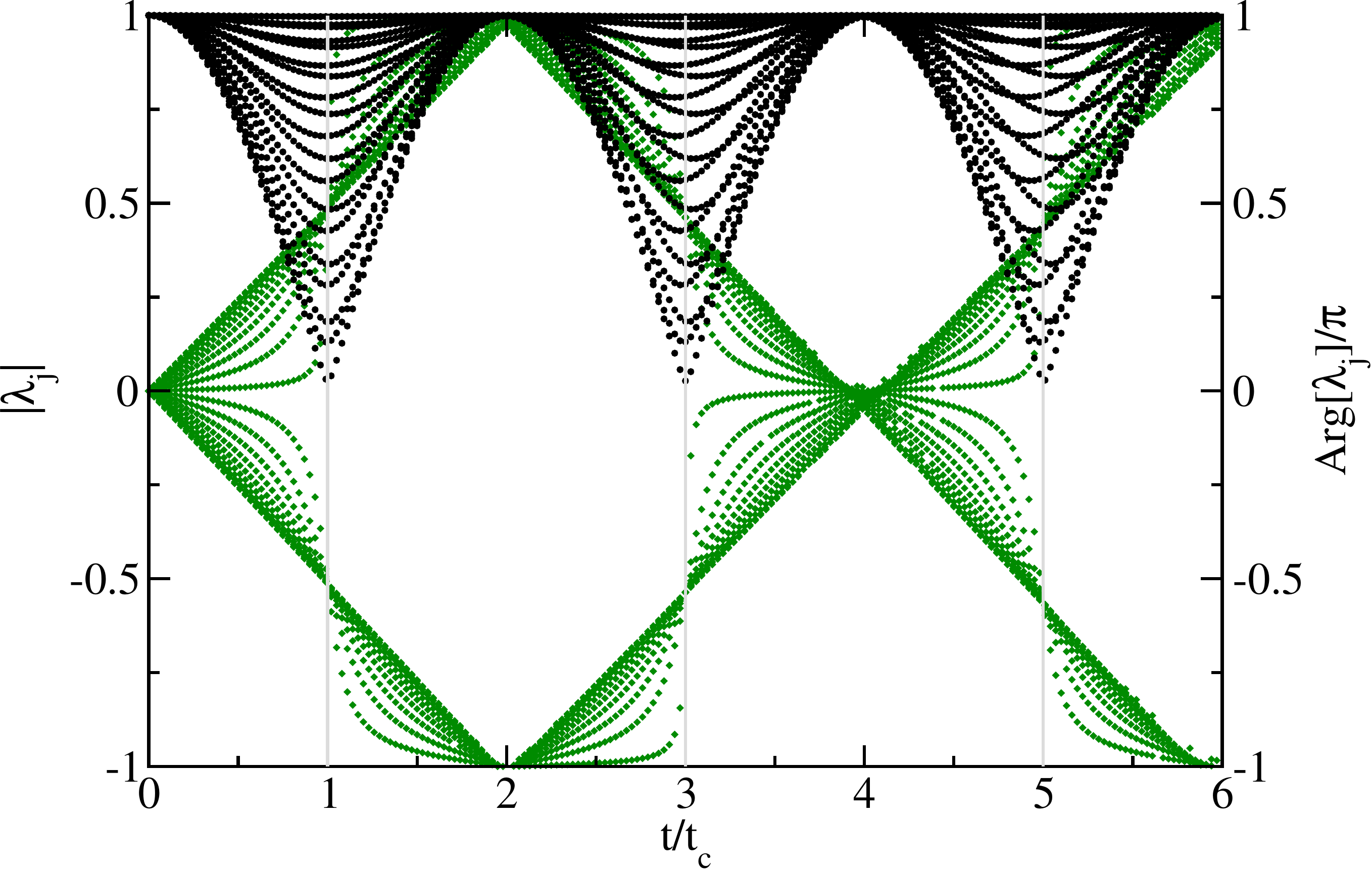}
\caption{Absolute value [black circles] and argument [green diamonds] of the eigenvalues of the matrix $\mathbf{M}$ for a symmetric quench at $|\delta|=0.95$ from the topological into the trivial phase for a system of size $N=80$.}
\label{Fig9}
\end{figure}

For the quench from the topological to the trivial phase we obtain a
slightly different picture. While at the critical times there is still
a pair of eigenvalues which approach zero, the eigenvalues no longer
remain close to zero for $t\in [t_c,3t_c]$, see Fig.~\ref{Fig9}.
The direction of the quench can thus clearly be distinguished from the
spectrum of $\mathbf{M}$. From the almost symmetric spectrum around
$t_c$ for the topological to trivial quench it is in particular clear
that the boundary contribution is of similar magnitude on both sides
of the transition consistent with the scaling results shown in
Fig.~\ref{Fig3}.

\begin{figure}[!ht]
\includegraphics*[width=0.99\linewidth]{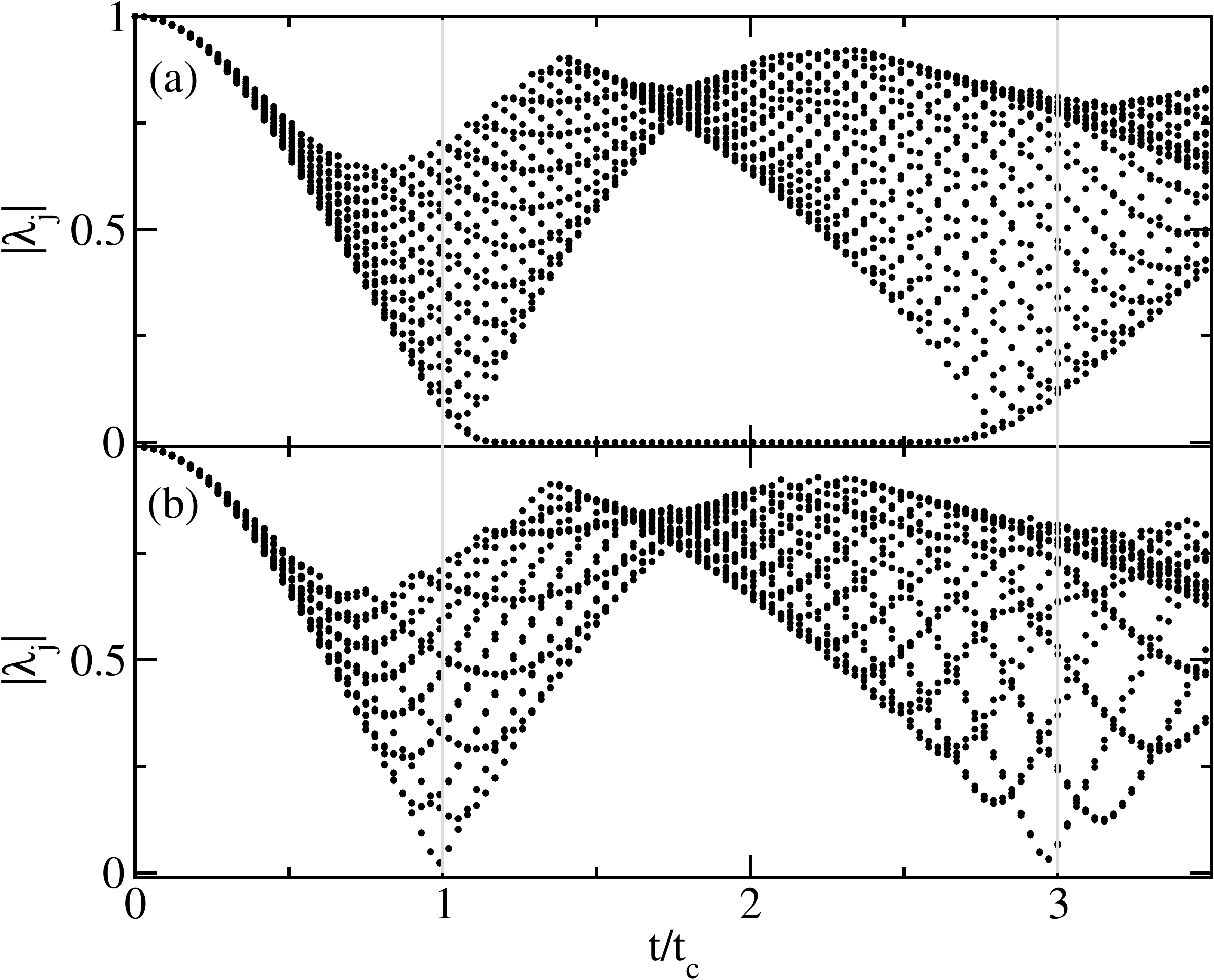}
\caption{Absolute value [black circles] of the smallest 20
eigenvalues of the matrix $\mathbf{M}$ for a symmetric quench at
$|\delta|=0.3$ from (a) the trivial to the topological phase, and (b)
vice versa, for a system of size $N=160$.}
\label{Fig10}
\end{figure}

So far we have investigated quenches for large dimerizations
$|\delta|\lesssim 1$ where the system is almost perfectly dimerized
and the Loschmidt spectrum is particularly simple. A natural question
to ask is whether or not the spectrum is still useful to detect if
edge states are present in the final Hamiltonian for smaller
dimerizations. To this end, we present in Fig.~\ref{Fig10} data for
symmetric quenches at $|\delta|=0.3$.
While the spectrum becomes more complex, the direction of the quench
is still obvious. In particular, a pair of eigenvalues close to zero
for $t\in [t_c,3t_c]$ persists for the trivial to topological quench.

\begin{figure}[!ht]
\includegraphics*[width=0.99\linewidth]{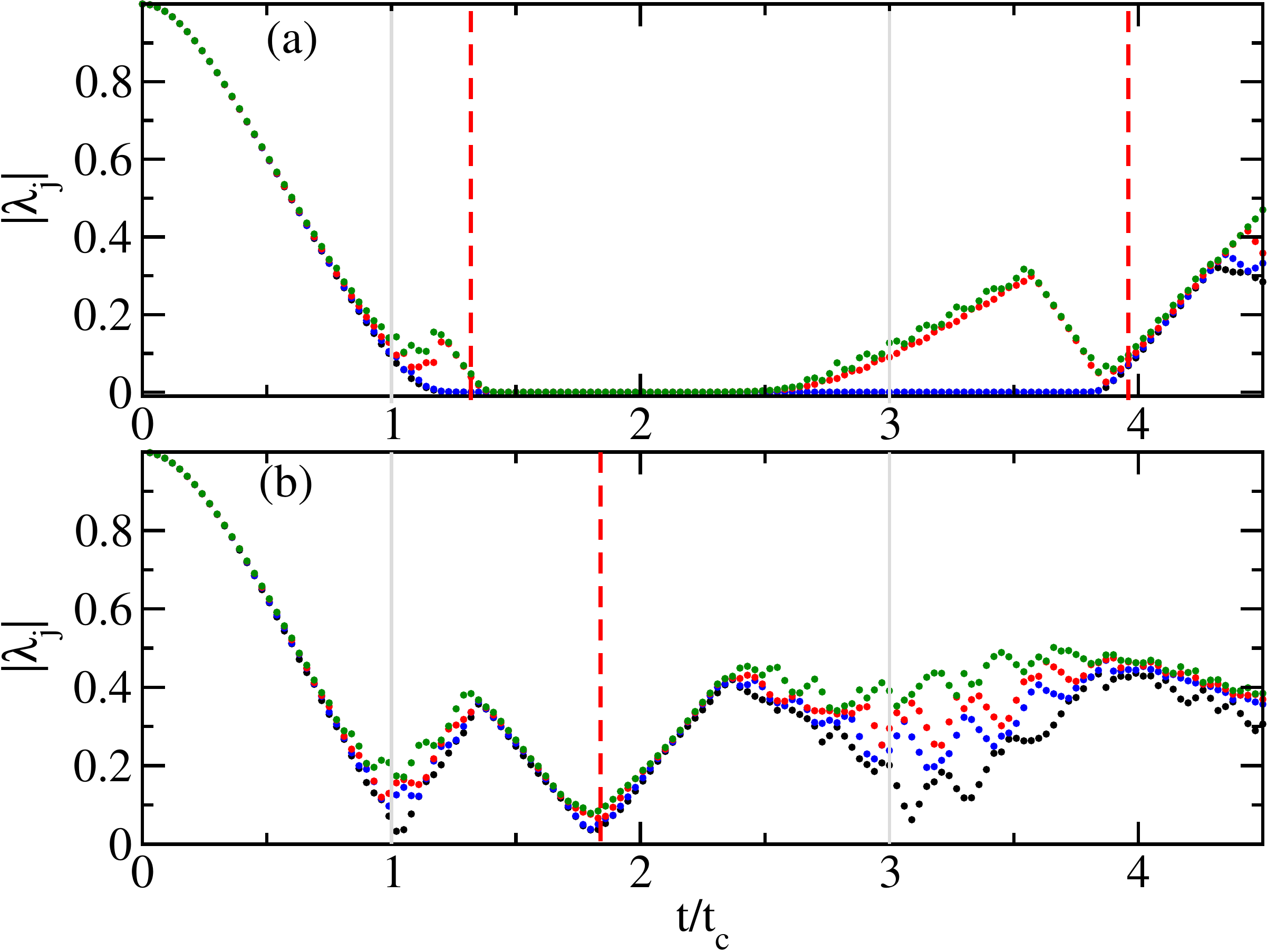}
\caption{Absolute value of the smallest four eigenvalues of the matrix $\mathbf{M}$ for the quenches in Fig.~\ref{Fig6}, for (a) $\nu_{\textrm{w}}=1$ to $\nu_{\textrm{w}}=3$, and (b) vice versa, for a system of size $N=600$. Eigenvalues are colored as an aid to the eye.}
\label{Fig11}
\end{figure}

In the long range Kitaev model where we are able to consider quenches
between more general winding numbers a similar structure is seen in
the spectrum of $\mathbf{M}$, see Fig.~\ref{Fig11}. Although finite
size effects are still present, it is already obvious that the results
are again very different for the two quench directions. Note that for
the quench from $\nu_{\textrm{w}}=1$ to $\nu_{\textrm{w}}=3$ the time
evolving Hamiltonian, $H_1$, has two additional pairs of edge states,
compared to the initial Hamiltonian, $H_0$.
For the quench where the number of edge states increases
[Fig.~\ref{Fig11}(a)] we find, in particular, an extended time
interval between critical times where $2$ or $4$ eigenvalues are
zero. In fact two eigenvalues are close to zero for $t\in[t_c,3t_c]$
and two different eigenvalues are close to zero for
$t\in[t'_c,3t'_c]$, where $t'_c$ is the larger of the two critical
times for this quench. For the opposite direction, on the other hand,
we find a roughly symmetric structure around the first two critical
times very similar to the SSH case. Note that finite size effects
strongly influence the results for $t/t_c\gtrsim 4$. Compatible with
the data for the SSH and the Kitaev chain is thus the idea that for
the boundary contribution $l_B(t)$ and for the Loschmidt spectrum,
$\text{spec}(\mathbf{M})$, it is important how many more or fewer edge
states are present for $H_1$ as compared to $H_0$.

\section{Long-range entanglement}
\label{Entangle}
We have established numerically a bulk-boundary correspondence and
could show that the spectrum of $\mathbf{M}$ contains information
about the edge states. However, the spectrum of this matrix is not an
easily measurable quantity and also does not provide a physical
picture about what happens to the edge states during the time
evolution. In this section we therefore want to connect the changes on
the Loschmidt echo with the dynamical entanglement properties of the
system. To this end, we consider the entanglement between the two
halves of an open chain with an even number of sites. The entanglement
entropy is defined as the von-Neumann entropy of a reduced density
matrix
\begin{equation}
 S_{\textrm{ent}}(t) =-\Tr \{\rho_A(t) \ln \rho_A(t)\}\,
\end{equation}
with $\rho_A(t)=\Tr_B |\Psi(t)\rangle\langle\Psi(t)|$ and
$|\Psi(t)\rangle = \text{e}^{-iH_1t}|\Psi_0\rangle$ being the
time-evolved state. The system has been divided up into two blocks $A$
and $B$ of equal size. Here we will focus on the SSH model.

For a Gaussian model the entanglement between a subsystem and the rest
can be calculated from the correlation matrix $\mathbf{\C}(t)$ defined
in Eq.~\eqref{Cij} with the now time-dependent two-point correlations
restricted to lattice sites within the subsystem.\cite{Peschel2003}
The entanglement entropy is then given by
\begin{equation}
\label{Sent}
S_{\textrm{ent}} = -\sum_j \left[ \eta_j\ln\eta_j +(1-\eta_j)\ln(1-\eta_j)\right] \, 
\end{equation}
with $\eta_j$ being the eigenvalues of $\mathbf{\C}(t)$.

\begin{figure}[!ht]
\includegraphics*[width=0.99\linewidth]{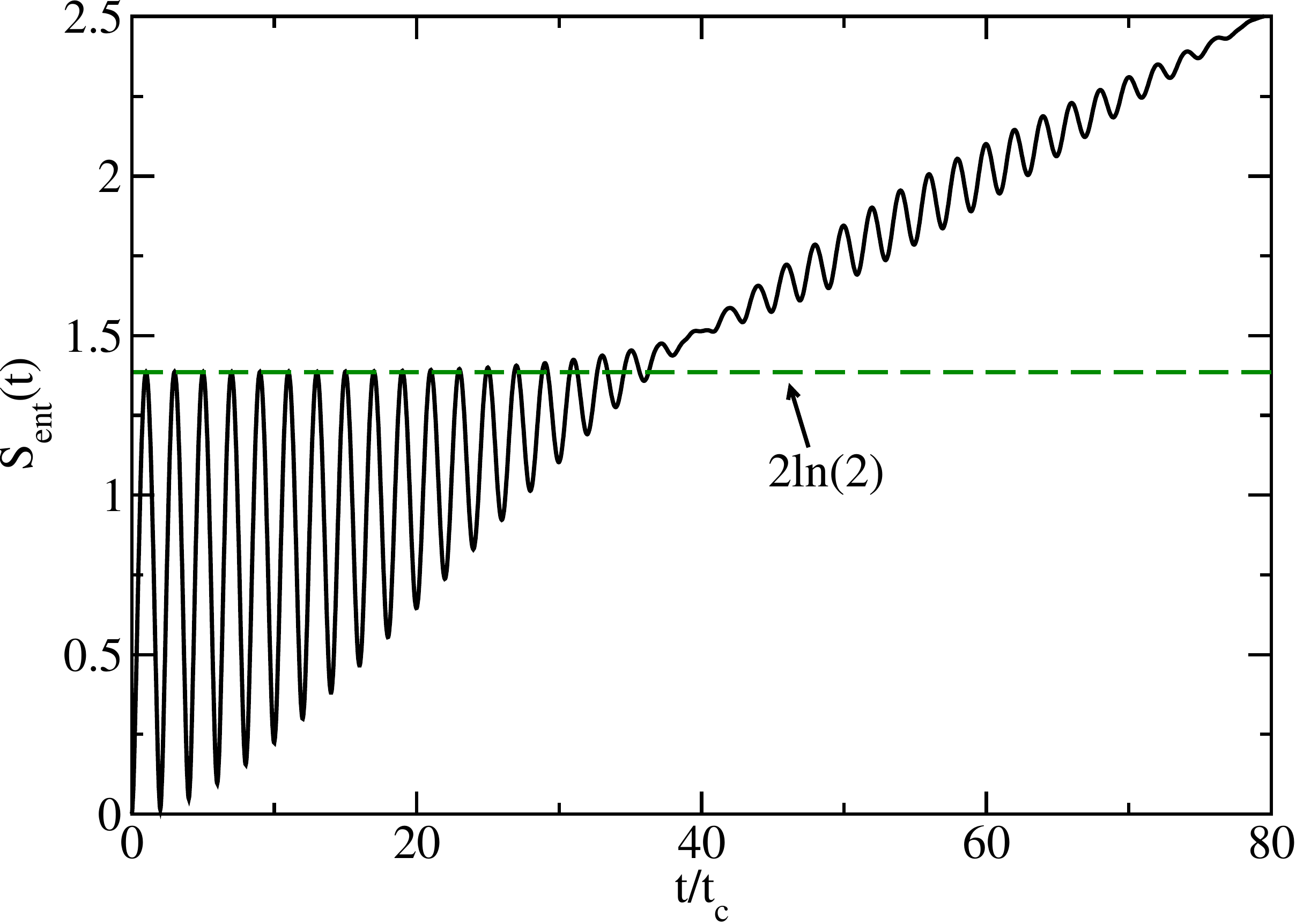}
\caption{$S_{\textrm{ent}}(t)$ between two halves of the system for a symmetric quench at $|\delta|=0.95$ from the trivial to the topological phase with $N=32$.}
\label{Fig12}
\end{figure}

\begin{figure}[!ht]
\includegraphics*[width=0.99\linewidth]{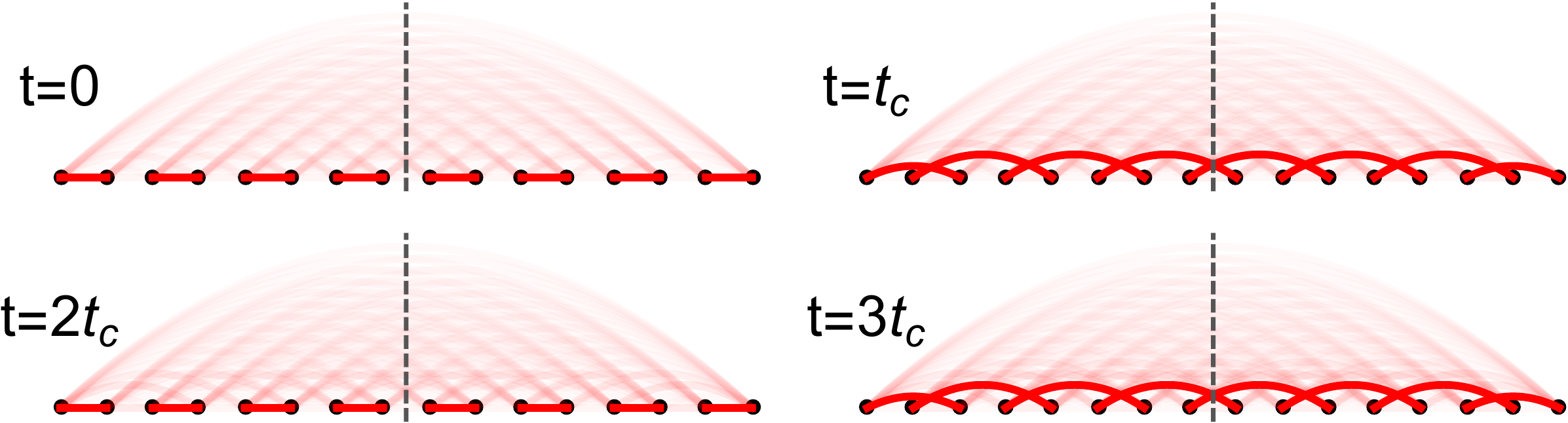}
\caption{Time evolution of two-point correlations for the same quench as in Fig.~\ref{Fig12} for a system with $N=16$ sites and times $t=0,t_c,2t_c,3t_c$. The opacity indicates the strength of the correlation and the dashed gray line is the cut between the two subsystems.}
\label{Fig13}
\end{figure}

In Fig.~\ref{Fig12} the entanglement entropy for a symmetric quench
from the trivial into the topological phase at large $|\delta|$ is
shown. 
$S_{\textrm{ent}}(t)$ is showing oscillations with local maxima
located exactly at the DPT's. For short times, in particular,
$S_{\textrm{ent}}(t_c)\approx 2\ln 2$ while at longer times the
entanglement entropy on average starts to increase linearly as is
expected for a global quench. These observatios can be explained as
follows: In the strongly dimerized case, each dimer bond between the
two subsystems is in a fully entangled state, $(|10\rangle \pm
|01\rangle)/\sqrt{2}$, and contributes $\ln 2$ to the entanglement
entropy. The data in Fig.~\ref{Fig12} therefore indicate that there
are two dimer bonds between the subsystems at the critical times
$(2n+1)t_c$. At times $2nt_c$, on the other hand, the two subsystems
are almost completely disentangled. This picture is confirmed by
directly considering the two-point correlations in the system as a
function of time, see Fig.~\ref{Fig13}.
The system starts in a topologically trivial, strongly dimerized state
$|\Psi_0\rangle=|\Psi(t=0)\rangle$ and thus
$S_{\textrm{ent}}(t=0)\approx 0$. At the critical time $t_c$, two
dimer bonds have formed which cross the cut between the subsystems
leading to a $S_{\textrm{ent}}(t_c)\approx 2\ln 2$. For short times,
the system oscillates between these two configurations with other
correlations slowly starting to build up and finally leading to a
linearly increasing entanglement entropy. Note that for a finite
system this linear increase will be cut off at
$S^{\textrm{max}}_{\textrm{ent}}=\frac{N}{2}\ln 2$.

\begin{figure}[!ht]
\includegraphics*[width=0.99\linewidth]{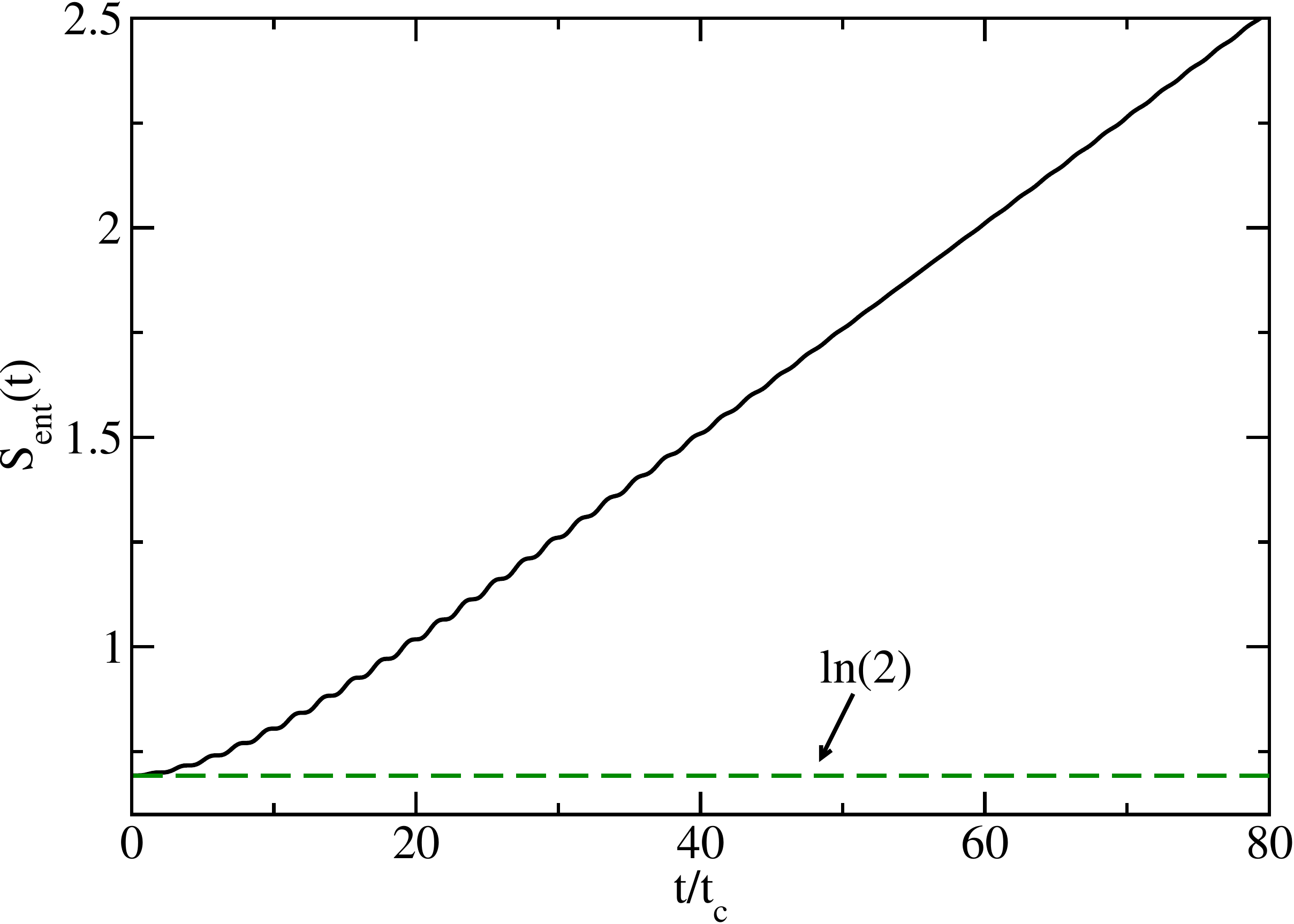}
\caption{$S_{\textrm{ent}}(t)$ between two halves of the system for a symmetric quench at $|\delta|=0.95$ from the topological to the trivial phase with $N=32$.}
\label{Fig14}
\end{figure}

\begin{figure}[!ht]
\includegraphics*[width=0.99\linewidth]{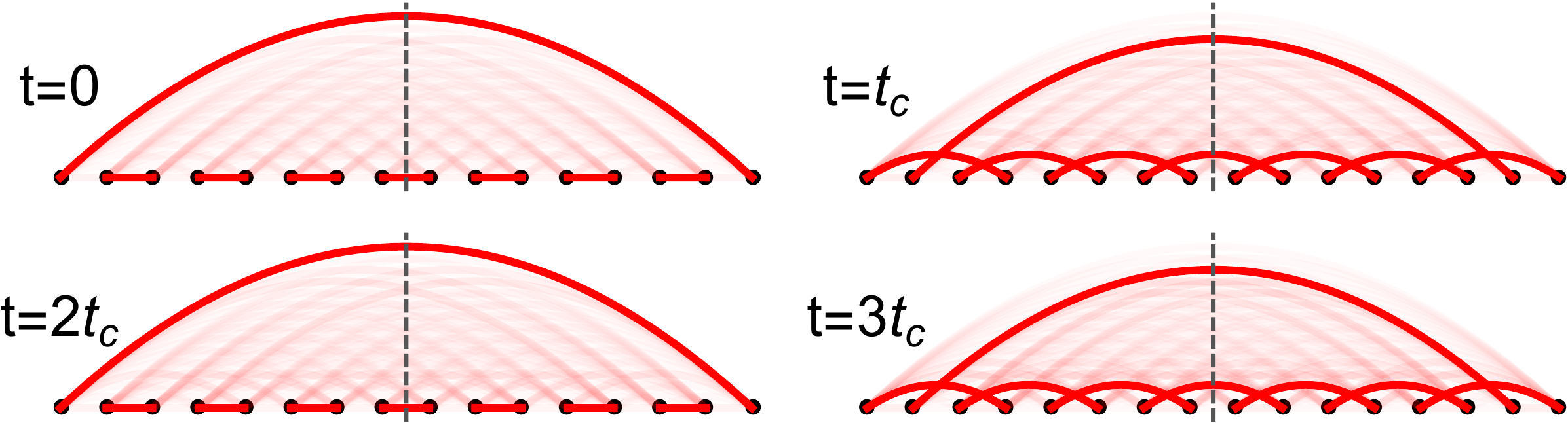}
\caption{Time evolution of two-point correlations for the same quench as in Fig.~\ref{Fig14} for a system with $N=16$ sites and times $t=0,t_c,2t_c,3t_c$. The opacity indicates the strength of the correlation and the dashed gray line is the cut between the two subsystems. Long-range entanglement between the edges of the system is present at all times.}
\label{Fig15}
\end{figure}

For the quench in the opposite direction, shown in Fig.~\ref{Fig14},
$S_{\textrm{ent}}$ also shows oscillations with a frequency set by the
critical time $t_c$.
In this case $S_{\textrm{ent}}(t=0)\approx 2\ln 2$ and for long times
the entanglement entropy shows again the expected linear increase. To
understand this behavior it is again instructive to consider the time
evolution of the two-point correlations, see Fig.~\ref{Fig15}.
At time $t=0$ a nearest-neighbor dimer and a dimer between the two
edge sites is cut. Such long-range entanglement between the boundaries
of an open SSH chain has been discussed previously in
Ref.~\onlinecite{CamposVenuti2007a}. Interestingly, the long-range
entanglement persists at all times during the unitary time evolution
despite the strong quench perturbing the system. While at times
$2nt_c$ the two edge sites are strongly entangled, the long-range
entanglement moves to the sites one removed from the edge for times
$(2n+1)t_c$, see Fig.~\ref{Fig15}. The different entanglement
structure in the two cases explains the oscillations seen in the
entanglement entropy shown in Fig.~\ref{Fig14} while the slow build-up
of additional correlations explains the linear increase of
$S_{\textrm{ent}}(t)$ at longer times.

\section{Conclusions}
\label{Concl}
In this paper we have studied dynamical phase transitions in open
chains with symmetry protected topological phases. Specifically, we
have concentrated on two examples: the SSH chain and a long-range
Kitaev model. In both cases we have shown that for a quench between
different topological phases there is not only a cusp in the bulk
return rate but also a jump in the boundary ($1/N$) contribution. In
contrast to the bulk part, the boundary return rate $l_B(t)$ is
sensitive to the direction of the quench.  For the SSH model, in
particular, we found that the jump in $l_B(t)$ at a DPT is orders of
magnitude larger for a quench from the trivial to the topological
phase than in the other direction.  A clear qualitative difference
between the quench directions can also be seen in the Loschmidt
eigenvalue spectrum: While for the quench into the topological phase
two eigenvalues which behave very differently on both sites of the DPT
are largely responsible for the boundary contribution, the spectrum
near a DPT is almost symmetric for a quench in the opposite direction.

The critical times $t_c$ at which DPT's occur are also clearly visible
as oscillations in the entanglement entropy between two halves of the
SSH chain. We found that these oscillations can be explained by the
time-dependent structure of two-point correlations. At short times and
large dimerizations the system oscillates between two configurations
of correlations. Starting from the topological phase we discussed, in
particular, the long-range entanglement which is transferred from the
boundary sites onto the sites one removed from the boundary going from
times $2nt_c$ to $(2n+1)t_c$. Quite surprisingly, the long-range
entanglement in this case remains stable despite the fact that the
quench introduces a large perturbation to the system.

\acknowledgments
JS acknowledges support by the Natural Sciences and Engineering
Research Council (NSERC, Canada) and by the Deutsche
Forschungsgemeinschaft (DFG) via Research Unit FOR 2316. Support for this research at Michigan State University (NS) was provided by the Institute for Mathematical and Theoretical Physics with funding from the office of the Vice President for Research and Graduate Studies.

\end{document}